\documentclass[aps,preprint]{revtex4}
\usepackage{amssymb,epsf}
\usepackage{amsmath}
\usepackage{graphicx}

\begin{document}
\title{Non-Abelian Lovelock-Born-Infeld Topological Black Holes}
\author{N. Bostani$^{1,2}$ \footnote{nbostani@ihep.ac.cn}and N. Farhangkhah$^3$ } \affiliation{$^1$ Key Laboratory of
Particle Astrophysics, Institute of High Energy Physics,
Chinese Academy of Sciences, Beijing 100049, China\\
$^2$Research Institute for Astrophysics and Astronomy of Maragha
(RIAAM), Maragha, Iran \\$^3$Physics Department and Biruni
Observatory, College of Sciences, Shiraz University, Shiraz 71454,
Iran}

\begin{abstract}

We present the asymptotically AdS solutions of the Einstein
gravity with hyperbolic horizons in the presence of
$So(n(n-1)/2-1, 1)$ Yang-Mills fields governed by the non-Abelian
Born-Infeld lagrangian. We investigate the properties of these
solutions as well as their asymptotic behavior in varies
dimensions. The properties of these kind of solutions are like the
Einstein-Yang-Mills solutions. But the differences seems to appear
in the role of the mass, charge and born-Infeld parameter $\beta$,
in the solutions. For example in Einstein-Yang-Mills theory the
solutions with non-negative mass cannot present an extreme black
hole while that of in Einstein-Yang-Mills-Born Infeld theory can.
Also the singularities in higher dimensional Einstein-Yang-Mills
theory for non-negative mass are always spacelike, while depending
on choosing the parameters, we can find timelike singularities in
the similar case of Einstein-Yang-Mills-Born-Infeld theory. We
also extend the solutions of Einstein to the case of Gauss-Bonnet
and Lovelock gravity. It is shown that, these solutions in the
limits of $\beta\rightarrow0$, and $\beta\rightarrow\infty$,
represent pure gravity and gravity coupled with Yang-Mills fields,
respectively.

\end{abstract}
 \maketitle

\section{Introduction}

Non-linear electrodynamics was proposed in the thirties to remove
singularities associated with charged pointlike particles. Among
the non-linear theories of electrodynamics Born-Infeld (BI) theory
\cite{BI} is distinguished, since BI type actions arise in many
different contexts in superstring theory \cite{Fradkin,Leigh}.
Also In the case of a D-brane, it has been known that the
effective action on the brane is Born-Infeld action if all
derivative terms are neglected \cite{Abo}. The effective action of
several D-branes is very important in the development of the
understanding of the non-perturbative superstring theory, such as
Matrix theory \cite{Banks}. Tseytlin \cite{Tse} proposed that if
all derivative terms are neglected, the effective action on the
branes is a non-Abelian generalization of the Born-Infeld action.
There are a lot of outstanding classical solutions to both the
Abelian and non-Abelian Born-Infeld theories. Born-Infeld black
holes in (A)dS spaces has been discussed in \cite{Cai1}. A type of
a particle-like solution in the non-Abelian Born-Infeld model was
obtained by Gal'tsov and Kerner (GK) \cite{Gal}. More close
relationship between these two particle-like configurations
becomes clear in \cite{Kunz,Gal2,HalilBI}. Among theories of
non-Abelian gauge fields, the Yang-Mills theory may be regarded as
the most fundamental one in elementary particle physics. The role
of the Yang-Mills field in gravity has become an interesting topic
of studies. Physical significance of the particle-like solutions
of Einstein-Yang-Mills (EYM) field equations found by Bartnik and
McKinnon (BK) \cite{BK} as well as their possible role in the
string-inspired models remains rather obscure. The attentions
towards the Einstein-Yang-Mills system even became more, after the
discovery of the first known example of hairy black holes
\cite{Vol1,Bizon} which are not uniquely characterized by their
conserved charges and so violate manifestly the no-hair conjecture
theory. In particular, particle-like, soliton-like and black hole
solutions in the combined Einstein-Yang-Mills (EYM) models in
different dimensions, shed new light on the complex features of
compact object in these models \cite{HalilE}. (See \cite{Vol} for
an overview).

With the advent of string theory, the possibility to have
extra-dimensions became one of the most promising possibility to
extend the standard model of particles physics. Higher dimensional
theories of gravity may present some new features which are absent
in four dimensions. Indeed in four dimensions the only
gravitational action that can be built from curvature invariants
leading to second order equations in the metric is the
Einstein-Hilbert action. The situation changes in higher
dimensions. In five dimensions one can add for example a
Gauss-Bonnet term to the action . In higher dimensions one can add
higher curvature powers to the action. Such higher curvature power
theories leading to second order equations for the metric are
known as Lovelock theories \cite{Lov}. There are lots of
considerable works in Lovelock gravity (for example see
\cite{Cai2}).

In this present Letter, we will study the topological black hole
of lovelock gravity in the present of non-abelian version of
Born-Infeld theory in (n+1)-dimensions with the gauge group
$SO(n(n-1)/2-1,1)$. The coupling of the two non-linear ,
Yang-Mills and Born-Infeld fields, is a candidate for effective
action for superstrings and D-branes, besides the fundamental
importance in the context of gravitation theory. To obtain an
exact solutions, we use a modified version of Wu-Yang ansatz
\cite{Wu} which was originally introduced in $N = 4$ field theory
and for the first time was applied by Yasskin \cite{Yasskin}.
Recently, this ansatz has been used to find the EYM and GBYM
solutions with negative cosmological constant in higher dimensions
\cite{BoE,BoGB}.\\The out line of this paper is as follows. In
Sec. II we introduce the field equations of Lovelock gravity in
the presence of the energy-momentum tensor of the coupling of
Yang-Mills and Born-Infeld fields. In section \ref{EE}, we obtain
the (n+1)-dimensional solutions of the Einstein gravity and
investigate their properties in varies dimensions. Section
\ref{Love4} devotes to the general solutions to the Gauss-Bonnet
and Lovelock gravity, and their properties respectively. We finish
our paper with some concluding remarks.

\section{Field Equations}

 As was shown by Lovelock in the
early seventies \cite{Lov}, the possible corrections to Einstein
gravity are quite limited, since the only symmetric, divergence
free tensor that can be constructed out of the metric and its
first two derivatives in a (n+1)- dimensional space-time is
\begin{equation}
{\cal G}_{\mu\nu}=\sum_{k=0}^{[n/2]} \alpha'_{k} \:{\cal
G}_{\mu\nu}^{(k)}=8\pi T_{\mu\nu} \label{fiel}
\end{equation}
\begin{equation}
{\cal G}^\mu \: _\nu^{(k)}=- \frac{1}{2^{k+1}}\:\delta_{\nu\: j_1
j_2....j_{2k}}^{\mu\: i_1 i_2....i_{2k}}\:R^{j_1 j_2}_{i_1
i_2}\:....R^{j_{2k-1} j_{2k}}_{i_{2k-1} i_{2k}}\notag\\
\end{equation}
Where $[d/2]$ denotes the integer part of $d/2$, $T_{\mu\nu}$ is
the energy-momentum tensor and $\alpha' _k$ are Lovelock constants
which represent the coupling of the terms in the whole Lagrangian
and give the proper dimensions. Usually in order the Einstein
gravity to be recovered in the low energy limit, the constant
$\alpha'_{0}$ should be identified as the cosmological constant up
to a constant, $\alpha'_{0}=-2\Lambda$, and $\alpha'_{1}$ should
be positive (for simplicity one may take $\alpha'_{1}=1$). As In
this context, we are interested in asymptotically anti-de Sitter
(AdS) solutions, we set $\Lambda =-n(n-1)/2l^{2}$.

We want to examine the Lovelock-Yang-Mills-Born-Infeld system for
a compact, semi-simple gauge group ${\mathcal G}$ with structure
constants $C_{bc}^{a}$. The basic elements of the model are
$({\cal
 M},g_{\mu\nu},A_{\mu }^{\left( a\right) })$, where ${\cal M}$ is the spacetime manifold with
metric $g_{\mu\nu}$ and $A_{\mu }^{\left( a\right) } $ are the
gauge potentials. The metric tensor of the gauge group is
\cite{BoE}
\begin{equation*}
\gamma _{ab}=-C_{ad}^{c}C_{bc}^{d}/\left| \det
C_{ad}^{c}C_{bc}^{d}\right| ^{1/M},
\end{equation*}
where $M$ is the dimension of the gauge group, the Latin indices
$a$, $b$.... go from $1$ to $M$, and the repeated indices is
understood to be summed over.

 In Eq.(\ref{fiel}), $T_{\mu\nu}$ is the energy momentum tensor given as
\begin{equation}
T_{\mu\nu } =\frac{1}{4\pi}\left(\frac{%
\gamma_{ab}}{\Gamma}{}F_{\mu \lambda }^{(a)}F_{\nu }^{(b)\lambda }+\frac{1}{4}g_{\mu\nu }{\cal L}_m\right)  \label{Field3}\\
\end{equation}
where therein
\begin{equation}
\Gamma = \sqrt{1+\frac{\gamma_{ab}}{2\beta ^{2}}\:{F}_{\lambda \sigma
}^{(a)}F^{(b)\lambda \sigma
}} \label{etta}\\
\end{equation}
and
\begin{equation}
F^{(a)}_{\mu\nu}=\nabla _{\mu }A_{\nu }^{(a)}-\nabla _{\nu }A_{\mu }^{(a)}+\frac{1}{2e}%
C_{bc}^{a}A_{\mu }^{(b)}A_{\nu }^{(c)}\label{ttt}
\end{equation}

In the equations (\ref{Field3}),(\ref{etta}) and (\ref{ttt}),
$\beta $ is the Born-Infeld parameter which has the dimension of $
length^{-2}$, $F_{\mu\nu}^{(a)}$'s are the Yang-Mills fields, $e$
is the coupling constant and ${\cal L}_m$, The matter term, is the
F(2) nonabelian and nonlinear action density
\begin{equation}
 \mathcal{L}_m=4\beta
^{2}\left(1-\Gamma\right). \label{Lag3}
\end{equation}

In the limit $\beta \rightarrow 0$, $\mathcal{L}_m$ becomes equal
to zero and in the limit $\beta \rightarrow \infty $ it reduces to
the standard Yang-Mills form, $\mathcal{L}_m=-\gamma_{ab}F_{\mu
\nu }^{(a)}F^{(b)\mu \nu }$, so we expect to obtain the
corresponding solutions in these both limits.

The YMBI gauge fields  $F_{\mu\nu}^{(a)}$'s satisfy the
nonlinear-nonabelian equation

\begin{equation}
\nabla _{\mu }(\frac{1}{\Gamma}F^{\mu \nu (a)})+\frac{1}{\Gamma e}%
C_{bc}^{a}A_{\mu }^{(b)}F^{(c)\mu \nu }=0, \label{BIField}
\end{equation}

We consider the metric of the following form:
\begin{equation}
ds^{2}=-f(r)\;dt^{2}+\frac{dr^{2}}{f(r)}+r^{2}d\;\Omega
_{n-1}^{2}\label{Metr1}
\end{equation}
where $d\Omega _{n-1}^{2}=d\theta
_{1}^{2}+\sinh\theta_{1}^{2}d\theta_{2}^{2}+\sinh\theta_{1}^{2}\sum_{i=3}^{n-1}\prod_{j=2}^{i-1}\sin\theta_{j}^{2}d\theta_{i}^{2}$,
is the line element of $(n-1)$- dimensional hypersurface with
constant negative curvature and volume $V_{n-1}$.

In order to obtain The gauge potential in higher dimensional
spacetimes we use the  (n+1)- dimensional Wu-Yang Ansatz \cite{Wu} as follows%
\begin{eqnarray*}
A^{(a)} &=&\frac{e}{r^{2}}\left( x_{i}dx_{n}-x_{n}dx_{i}\right)
;\text{ \ \
\ \ }i=1...n-1, \\
A^{(b)} &=&\frac{e}{r^{2}}\left( x_{i}dx_{j}-x_{j}dx_{i}\right)
;\text{ \ \ \ \ \ \ }i<j,
\end{eqnarray*}
where $a$ and $b$ run from $1$ to $n-1$ and $n$ to $n(n-1)/2$,
respectively and
\begin{eqnarray*}
x_{1} &=&r\sinh \theta_{1} \prod_{j=2}^{n-1}\sin \theta _{j}, \\
x_{i} &=&r\sinh \theta_{1} \cos \theta
_{n-i+1}\prod_{j=2}^{n-i}\sin \theta
_{j};\text{ \ \ }i=2...n-1, \\
x_{n} &=&r\cosh \theta_{1} ,
\end{eqnarray*}
The Lie algebra of this gauge group is $SO(n(n-1)/2-1,1)$ with the
metric tensor of $\gamma _{ab}=\epsilon _{a}\delta _{ab}$, which
there is no sum on $a$ and
\begin{equation*}
\epsilon _{a}=\left\{
\begin{array}{ll}
-1 & \ \ \ 1\leq a\leq n-1 \\
1 & \ \ \ n\leq a\leq \frac{n(n-1)}{2}
\end{array}
\right.  \label{kappa}
\end{equation*}

Substituting Eqs. (\ref{Metr1}) and (\ref{ttt}) in (\ref{etta}),
we have
\begin{equation}
\Gamma(r)=\sqrt{1+\frac{%
(n-1)(n-2)e^{2}}{2\beta^2r^{^{ }4}}}\label{eta}
\end{equation}

\section{Static Solutions in
Einstein Gravity}\label{EE}

In this section the goal is to find the exact solutions of
Einstein gravity in the presence of YMBI fields. In this case, the
field equation (\ref{fiel}) becomes
\begin{equation}
{\cal G}_{\mu \nu }^{(1)}-\frac{n(n-1)}{2l^2} g_{\mu \nu }=8\pi
T_{\mu \nu }. \label{FieldEin}
\end{equation}

Where ${\cal G}_{\mu \nu }^{(1)}$ is just the Einstein tensor. To
find the function $f(r)$, one may use any components of Eq.
(\ref{FieldEin}). The tt-component of the above equation using (%
\ref{Field3}) and (\ref{eta}) is
\begin{equation}
\left[ r^{n-2}(1+f)\right] ^{\prime }-\frac{n}{l^{2}}%
r^{n-1}-\frac{4\beta^2}{n-1}r^{n-1}\left(1-\Gamma\right)=0
\label{EqttE}
\end{equation}
For $N=n+1=5$ the solution is obtained to be:
\begin{equation*}
f(r)=-1-\frac{3m}{r^{2}}+\frac{r^2}{l^2}-\frac{\beta^2r^2}{3}\left(1-\Gamma_4\right)-\frac{2e^2\ln(r)}{r^2}-
\frac{e^2\ln\left(1+\Gamma_4\right)}{r^2}, \label{F5E}
\end{equation*}

Where $\Gamma_4=\Gamma\:(n=4)$. For $N=n+1>5$\ the solution of the
field equation (\ref{EqttE}) is :
\begin{equation}
f(r)=-1-\frac{(n-1)m}{r^{n-2}}+\frac{r^2}{l^2}+\frac{4\beta^2r^2}{n(n-1)}\left[1-\digamma
(r)\right],\label{FrN}
\end{equation}

$\digamma (r)$ is defined as:

\begin{equation}
\digamma (r)=\frac{n}{r^{n}}\int r^{n-1}\Gamma(r)\: dr={_{2}F_{1}}\left( \left[ -{\frac{%
1}{2},-\frac{{n}}{{4}}}\right] ,\left[ 1-{\frac{{n}}{{4}}}\right]
,\Gamma^2-1 \right)   \label{SSS}
\end{equation}

Where $_{2}F_{1}([a,b],[c],z)$ is hypergeometric function. It is
apparent that\ in the limit $\beta \rightarrow 0$ Eq. (\ref{FrN})
will be
\begin{equation}
f(r)=-1-\frac{(n-1)m}{r^{2}}+\frac{r^2}{l^2}  \nonumber\label{F0E}
\end{equation}
which is the AdS solution to the Einstein equation. Using the fact
that $_{2}F_{1}([a,b],[c],z)$ has a convergent series expansion for $|z|<1$%
, we can find the limit of the $f(r)$ for large $r$\ and $\beta $
as
\begin{equation}
f(r)\rightarrow -1-\frac{(n-1)m}{r^{n-2}}+\frac{r^{2}}{l^2}-\frac{%
(n-2)e^{2}}{(n-4)r^{2}}  \label{finf}
\end{equation}
As expected this is the solution to the Einstein-Yang-Mills
equations introduced in \cite{BoE}.

The global structure of the spacetime is characterized by
properties of the singularities and horizons. It is easy \ to show
that the Kretschmann scalar $R_{\mu \nu \lambda \kappa }R^{\mu \nu
\lambda \kappa }$ diverges at $r=0$ and is finite everywhere else,
so $r=0$ is an essential singularity. The behavior of the solution
$f(r)$ \ at infinity\ is dominated by the term
$r^2\left(1/l^2+4\beta^2/n(n-1)\right)$ and so one can see that
$f(r)\rightarrow +\infty $ \ for large r's.  Also one can see that
the function $f(r)$ has a negative or
positive value near $r=0$ depending on the value of $-\frac{(n-1)m}{r^{n-2}}-%
\frac{4\beta^2r^2}{n(n-1)}\digamma (r)$ as this is the dominating
term
in $f(r)$ near $r=0.$ The hypergeometric function ${_{2}F_{1}}\left( \left[ -%
{\frac{1}{2},-\frac{{n}}{{4}}}\right] ,\left[
1-{\frac{{n}}{{4}}}\right] ,\Gamma^2-1 \right) $, has different
forms in different dimensions. \

By considering an integer number $a$, we obtain the explicit form
of the hypergeometric function as follows.

\bigskip \subsection{$\mathbf{n=8a,a=1,2,..}$}

\bigskip For these values of $n,$ as in ${_{2}F_{1}}\left( \left[ a,b\right] ,\left[ {c}\right] ,z \right)
,$ $\ c$ is always an integer, the integral (\ref{SSS}), is
expected to include a logarithmic term. As a case for $n=8$ the
metric function $f(r)$ is
\begin{equation*}
f=-1-\frac{7m}{r^{6}}+\frac{r^2}{l^2}+\frac{\beta
^{2}r^2}{14}\left(1-\Gamma_{8}\right)-\frac{3e^2}{4r^2}\Gamma_{8}\left[1-\ln\left(r^2+r^2\Gamma_{8}\right)^{\frac{\Gamma_8^2-1}{\Gamma_8}}\right]
\end{equation*}

Where $\Gamma_{8}=\Gamma(n=8)$. Of course one should note that
this solution in $\beta \rightarrow +\infty ,$ reduces to give the
relation (\ref{finf}) as expected, and the logarithmic term
vanishes in this case. But other than this case, this logarithmic
case plays a significant term in the properties of the
Yang-Mills-Born-Infeld solutions.

To see these properties we obtain the limit of $f(r)$ near $r=0$
for these values of \ $n=8a$. In this case we obtain:

\begin{equation*}
f\rightarrow -\frac{(8a-1)}{r^{8a-2}}\left[m-\lambda e^{4a}\ln
(\varsigma)\right],
\end{equation*}
Where
\begin{eqnarray*}
\lambda&=&\frac{(8a-1)^{2a-2}(4a-1)^{2a-1}}{2^{4a-1}\beta
^{4a-2}}\left(_{\phantom{d}{2a}}^{4a-1}\right) \\\varsigma
&=&\frac{(8a-1)(4a-1)e^{2}}{\beta^2}
\end{eqnarray*}

When $m>\lambda e^{4a}\ln
(\varsigma)$, $f(r)$ is negative and as $f(r)\rightarrow +\infty $, when r$%
\rightarrow +\infty $, so $f(r)$ has certainly one real
root and a black hole with one horizon exists. For non-negative mass if $%
\varsigma>1$ and  $m<\lambda e^{4a}\ln (\varsigma)$, $f(r)$ is
positive near $r=0$ and the solution may present an extreme black
hole, a black hole with two horizons or a spacetime without a
horizon. This is a property that does not happen in the
Einstein-Yang-Mills theory as the spacetime always presents naked
singularity for non-negative mass.

Also for non-negative mass of this case the singularity at $r=0$
is timelike, but in the Einstein-Yang-Mills theory the singularity
for the dimensions higher than five, is always spacelike.

for negative mass, the possibility to have spacelike singularity
exists if $\varsigma<1$ and $\mid $ $m\mid <\mid \lambda e^{4a}\ln
(\varsigma)\mid $, while in EYM gravity the singularity for
negative mass is timelike.

\subsection{ $\mathbf{n=4\:(2a+1)}$,\: a= 1,2,\:...}

In this case, the hypergeometric function again includes a logarithmic term.
For the behavior of the metric function near $r=0$ we have
\begin{equation*}
f\rightarrow -\frac{(8a+3)}{r^{8a+2}}\left[m+\lambda e^{4a+2}\ln
(\varsigma)\right],
\end{equation*}
Where in this case
\begin{eqnarray*}
\lambda&=&\frac{(8a+3)^{2a-1}(4a+1)^{2a}}{2^{4a+1}\beta
^{4a}}\left(_{\phantom{d}{2a}}^{4a+1}\right) \\\varsigma
&=&\frac{(8a+3)(4a+1)e^{2}}{\beta^2}
\end{eqnarray*}

By choosing proper parameters, this term can be negative or positive. For
negative mass this term can be positive if $\varsigma<1$, or if $%
\varsigma>1$ and also $\mid $ $m\mid >\mid \lambda e^{4a+2}\ln
(\varsigma)\mid .$ In this case extreme black hole exists for the
proper choose of the parameters. $\ $For non-negative mass also if
$\beta $ be sufficiently small and $\varsigma<1$, $f(r)$ will be
positive and the solution may present a black hole with one or two
horizons or a spacetime without a horizon. This is also the case
that does not happen in Yang-Mills theory without Born-Infeld or
in Maxwell-Born-Infeld theory.

\bigskip \subsection{$\mathbf{n=2a+1,a= 1,2,\:...}$}
In this case
\begin{equation*}
 f\rightarrow -2a\frac{m}{r^{2a-1}}
\end{equation*}
 This is similar to the case that happens in
Maxwell-Born-Infeld theory and just when $m<0,$ extreme black hole
may exists and for non-negative mass spacetime always presents
naked singularity.

\bigskip \subsection{ $\mathbf{n=2\:(4a\pm 1), a= 1,2,\:...}$}

In these two cases, the integral in the relation (\ref{SSS}) can
be solved easily. For example for $n=6$ and $n=10$ respectively,
solutions are obtained as

\begin{equation*}
f_{n=6}=-1-\frac{5m}{r^{4}}+\frac{r^2}{l^2}+\frac{2\beta
^{2}r^2}{15}\left(1-\Gamma_{6}\right)-\frac{4e^2}{3r^2}\Gamma_{6},
\end{equation*}

\begin{equation*}
f_{n=10}=-1-\frac{9m}{r^{8}}+\frac{r^2}{l^2}+\frac{2\beta
^{2}r^2}{45}\left(1-\Gamma_{10}\right)-\frac{8e^2}{15r^2}\Gamma_{10}(3-2\Gamma_{10}^2),
\end{equation*}

Where $\Gamma_{6}$ and $\Gamma_{10}$ are the amount of $\Gamma$
for $n=6$ and $n=10$ respectively. The behavior of $f(r)$ near
$r=0$, for the case $n=2(4a-1)$ is

\begin{equation*}
f\rightarrow -\frac{(8a-3)}{r^{8a-4}}\left[m+\frac{\lambda
}{\beta^{4a-3}} e^{4a-1}\right],
\end{equation*}
\begin{equation*}
\lambda=\frac{2^{6a-\frac{7}{2}}{(2a-1)}^{2a-\frac{3}{2}}{(8a-3)}^{2a-\frac{5}{2}}}{(4a-1)\left(_{\phantom{d}{2a-1}}^{4a-3}\right)}
\end{equation*}
While that of For $n=2(4a+1)$ is as follows

\begin{equation*}
f\rightarrow -\frac{(8a+1)}{r^{8a}}\left[m-\frac{\lambda
}{\beta^{4a-1}} e^{4a+1}\right],
\end{equation*}
\begin{equation*}
\lambda=\frac{2^{8a-1}a^{2a-\frac{1}{2}}\left(8a+1\right)^{2a-\frac{3}{2}}}{(4a+1)\left(_{\phantom{d}{2a}}^{4a-1}\right)}
\end{equation*}

So depending on proper choosing of the parameters $m$ and $e,$
black hole with one or two horizons, an extreme black hole and a
naked singularity may exist. Also in these two cases a spacelike
or timelike singularities may exist depending on the values of $m$
and $e$ and $\beta$.

We know a horizon is a null hypersurface defined by $r=r_{h}$ such that $%
f(r_{h})=0$ with finite curvatures, where $r_{h}$ is a constant
horizon radius. For all the cases that $f(r)$ is positive near
$r=0$, the extreme black hole may exists which therein, both
$f(r)$ and $f^{\prime }(r)$ are zero on the horizon radius
$r=r_{\mathrm{ext}}$ and can be calculated from (\ref{EqttE}) to
be
\begin{eqnarray*}
r_{\mathrm{ext}}&=&\frac{\sqrt{\frac{n-2}{n}}l}{1+2\zeta}\left\{1+\zeta\left(1+\sqrt{1+\frac{2ne^2}{(n-2)l^2}\left(2+\frac{1}{\zeta}\right)}\right)\right\}\\
\zeta&=&\frac{4\beta^2l^2}{n(n-1)}
\end{eqnarray*}

For these cases the spacetime of Eqs. (\ref{Metr1}) and
(\ref{FrN}) presents a naked singularity if $m<m_{\mathrm{ext}}$,
an extreme black hole for $m=m_{\mathrm{ext}}$ and a black hole
with two horizons provided $m>m_{\mathrm{ext}}$, where
$m_{\mathrm{ext}}$ is
\begin{equation}
m_{\mathrm{ext}}=-r_{\mathrm{ext}}^{n-2}\left(\frac{1}{(n-1)}+\frac{r_{\mathrm{ext}}^2}{(n-1)l^2}+\frac{4\beta^2r_{\mathrm{ext}}^2}{n{(n-1)}^2}\left[1-\digamma
(r_{\mathrm{ext}})\right]\right),\nonumber\label{FRN}
\end{equation}
The Hawking temperature is given by
\begin{equation}
T =\frac{f'(r_+)}{4\pi}=\frac{(n-2)}{r_+}\Upsilon(r_+)
\end{equation}
Where $r_+$ is the largest real root of $f(r)$ and
\begin{equation}
\Upsilon(r_+) =-1+\frac{nr_+^2}{(n-2)l^2}
+\frac{4\beta^2r_+^2}{(n-1)(n-2)}\left[1-\digamma(r_+)-\frac{r_+}{n}\digamma'(r_+)\right],\label{Gam}
\end{equation}
$\digamma'(r_+)$ is the value of the first derivative of
$\digamma(r_+)$ at $r = r_+$. It's notable that $T$ vanishes for
$m=m_{\mathrm{ext}}$.

\section{Static Solutions in Lovelock Gravity}\label{Love4}

The Gauss-Bonnet-Yang-mills field equation in the presence of YMBI
fields may be written as
\begin{equation}
{\cal G}_{\mu \nu }^{(1)}+\alpha'_2{\cal G}_{\mu \nu }^{(2)}-%
\frac{n(n-1)}{2l^2} g_{\mu \nu }=8\pi T_{\mu \nu },
\label{FieldGB}
\end{equation}
Where ${\cal G}_{\mu \nu }^{(2)}$ is the second order Lovelock
tensor introduced in \cite{Hoi}. The tt-component of the above
field equation for the metric (\ref{Metr1}) is:
\begin{eqnarray}
&&\left[r^{3}-2\alpha_2 r(f+1)\right]f'-(n-4)\alpha_2 (f+1)^{2}+(n-2)r^{ 2}(f+1) \notag \\
&&-\frac{4\beta^2}{(n-1)}r^4\left(1-\Gamma\right)-\frac{n}{l^2}r^4
=0 \label{EqttEG}
\end{eqnarray}

Where $\alpha_2=(n - 2)(n - 3)\alpha'_2$. It is a matter of
calculation to show that the solution of the field equation
(\ref{EqttEG}), for $N=n+1=5$ may be written as
\begin{equation*}
f(r)=-1+\frac{r^{ 2}}{2\alpha_2
}\left(1\mp\sqrt{1-\frac{4\alpha_2}{l^2}+\frac{12\alpha_2
m}{r^4}+\frac{4\alpha_2\beta^2}{3}\left(1-\Gamma_{4}\right)+\frac{4\alpha_2e^2\ln\left(r^2+r^2\Gamma_{4}\right)}{r^4}}\right)
\label{fGB,5}
\end{equation*}
Where $\Gamma_{4}=\Gamma\:(n=4)$. Solving the equation
(\ref{EqttEG}) for $N\geq 5$, we get
\begin{equation}
f(r)=-1+\frac{r^{ 2}}{2\alpha_2
}\left(1\mp\sqrt{1-\frac{4\alpha_2}{l^2}+\frac{4(n-1)\alpha_2
m}{r^n}-\frac{16\alpha_2\beta^2}{n(n-1)}\left[1-\digamma
(r)\right]}\right) \label{fGB,n}
\end{equation}

Where $\digamma (r)$ was introduced in (\ref{SSS}). There are two
families of solutions which correspond to the sign in front of the
square root in Eq. (\ref{fGB,n}). We call the family with minus
(plus) sign the minus (plus)-branch solution. The minus-branch
solution reduces to the solution in the
Einstein-Yang-Mills-Born-Infeld solution in the limit of $\alpha
\rightarrow 0.$ On the other hand, $f(r)$ diverges for the
plus-branch solution in this limit, and there is no counterpart in
the Einstein theory.

We notice that $f(r)$ takes the form
\begin{equation}
f(r)=-1+\frac{r^{2}}{2\alpha_2 }\left[1\pm \sqrt{1-\frac{4\alpha_2}{l^2}+\frac{%
4(n-1)\alpha_2 m}{r^{n}}}\right] \nonumber\label{fGBv}
\end{equation}
as $\beta \rightarrow 0$. \ Also in the limit $\beta \rightarrow \infty $ it
takes the form
\begin{equation}
f(r)=-1+\frac{r^{ 2}}{2\alpha_2 }\left[1\pm \sqrt{1-\frac{4\alpha_2}{l^2}+\frac{%
4(n-1)\alpha_2 m}{r^{ n}}+\frac{4(n-2)\alpha_2 e^{2}}{(n-4)r^{
4}}}\right] \nonumber\label{fGB}
\end{equation}

This is the solution to the Gauss-Bonnet-Yang-Mills equation
obtained in \cite{BoGB}\ as expected.

Of course one may note that $f(r)$ is imaginary for $r<r_{0} $ and
real for $r>r_{0}$\ where $r_{0}$ is the largest real root of the
following equation:
\begin{equation*}
m+\frac{r^n}{(n-1)}\left(\frac{1}{4\alpha_2}-\frac{1}{l^2}-\frac{4\beta^2}{n(n-1)}\left[1-\digamma
(r)\right]\right)=0
\end{equation*}

Thus one cannot extend the spacetime to the region $r<r_{0}.$ To
get rid of this incorrect extension we introduce the new radial
coordinate $\rho$ as
\begin{equation*}
\rho^{2}=r^{2}-r_{0}^{2}\Rightarrow dr^{2}=\frac{\rho^{2}}{%
\rho^{2}+r_{0}^{2}}d\rho^{2}.  \label{metric2}
\end{equation*}

With this new coordinate the metric (\ref{Metr1}) becomes
\begin{equation}
ds^{2}=-f(\rho)dt^{2}+\frac{\rho^{2}}{\rho^{2}+r_{0}^{2}}\frac{d\rho^{2}}{f(\rho)}%
+(\rho^{2}+r_{0}^{2})d\Omega ^{2}  \label{metric 2}
\end{equation}
Where now one should substitute $r=\sqrt{\rho^2+r_{0}^{2}}$ in Eq.
(\ref{fGB,n}). The new metric function has a singularity at
$\rho=0$ $(r=r_0)$.

Like in Einstein gravity, In Gauss-Bonnet gravity, black hole with
one or two horizons, an extreme black hole and a naked singularity
may exist. here, we just consider the condition of having extreme
black holes, for which the temperature vanishes. The Hawking
temperature can be obtained as
\begin{equation}
T=\frac{(n-2)r_+}{(r_+^2-2\alpha)}\left(
\Upsilon(r_+)+\frac{(n-4)\alpha}{(n-2)r_+^2}
\right)\\
\end{equation}
Where $\Gamma(r_+)$ was given in Eq. (\ref{Gam}) and $r_+$ is the
radius of outer horizon. It is a matter of calculation to show
that $m=m_{\mathrm{ext}}$,
\begin{equation*}
m_{\mathrm{ext}}=\frac{2r_+^{n-2}}{n(n-1)}\left(-1+
\frac{2\alpha_{2}}{r_+^2}+\frac{2r_+^3\beta^2}{n(n-1)}\digamma'
(r_+)\right)
\end{equation*}
is the solution of $T=0$.

To find explicitly the solutions to the Yang-Mills-Born-Infeld
equation in third order Lovelock gravity in the presence of
cosmological constant, we have to consider the third order
Lovelock-Yang-mills field equation as
\begin{equation}
{\cal G}_{\mu \nu }^{(1)}+\alpha _{2}^{\prime }{\cal G}_{\mu \nu
}^{(2)}+\alpha _{3}^{\prime }{\cal G}_{\mu \nu
}^{(3)}-\frac{n(n-1)}{2l^2} g_{\mu \nu }=8\pi T_{\mu \nu },
\label{FieldLove}
\end{equation}
Where now ${\cal G}_{\mu \nu }^{(3)}$ is the third order Lovelock
tensor introduced in \cite{Hoi}. The tt-component of the field
equation (\ref{FieldLove}) for the metric ansatz (\ref{Metr1}) is
derived to be:

\begin{eqnarray}
&&\left[ \alpha_{3}r (f+1)^{2} -2 \alpha_{2}r^{3}
(f+1)+r^{5}\right]
f^{\prime } +\frac{n-6}{3}\alpha_{3}(f+1)^{3}-(n-4)\alpha_{2}r^{2}(f+1)^{2},  \nonumber \\
&&+(n-2)r^{4}(f+1)-
\frac{n}{l^{2}}r^{6}-\frac{%
4\beta ^{2}r^{6}}{(n-1)}(1-\Gamma)=0,  \label{Eqf1}
\end{eqnarray}

\bigskip where prime denotes the derivative with respect to $r$ and we
define $\alpha _{2}^{^{\prime }}=\alpha _{2}/(n-2)(n-3)$ and $\alpha
_{3}^{^{\prime }}=\alpha _{3}/3(n-2)(n-3)(n-4)(n-5)$ for simplicity.

First we solve the equation (\ref{Eqf1}) for the case $%
N=n+1=7.$ The equation in this case admits the solution:

\begin{equation}
f(r)=-1+\frac{\alpha_2 r ^{2}}{\alpha_{3}}\left\{1+
\left(\sqrt{\gamma +j^{2}(r)}+j(r)\right) ^{1/3}-\gamma
^{1/3}\left( \sqrt{\gamma +j^{2}(r)}+j(r)\right) ^{-1/3}\right\}
 , \label{f1}
\end{equation}
where
\begin{eqnarray*}
\gamma &=&(\frac{\alpha_{3}}{\alpha_{2}^{2}}-1)^{3},\label{gam1}\nonumber\\
j(r) &=&1-\frac{3\alpha_3}{2\alpha^2_2}
+\frac{3\alpha_{3}^{2}}{2\alpha_2^3}\left(\frac{1}{l^2}-\frac{5m}{r^{6}}+\frac{2\beta
^{2}}{15}\left(1-\Gamma_6\right)-\frac{4e^2}{3r^4}\Gamma_6\right).
\label{jrho} \label{krho}
\end{eqnarray*}

Where $\Gamma_{6}$, as was mentioned before, is the amount of
$\Gamma$, for $n=6$. Also for $N=n+1>7$ the generalized solution
is the same as the relation (\ref{f1}) with $j(r)$ being:
\begin{equation}
j(r) =1-\frac{3\alpha_3}{2\alpha^2_2}
+\frac{3\alpha_{3}^{2}}{2\alpha_2^3}\left(\frac{1}{l^2}-\frac{(n-1)m}{r^{n}}+\frac{4\beta^2}{n(n-1)}\left[1-\digamma
(r)\right]\right)  \label{kn}
\end{equation}

\bigskip Now for this solution as $\beta \rightarrow 0,$ $k(r)$\ reduces to
\begin{equation*}
j(r)=1-\frac{3\alpha_3}{2\alpha^2_2}
+\frac{3\alpha_{3}^{2}}{2l^2\alpha_2^3}-\frac{3(n-1)\alpha_{3}^{2}m}{2\alpha_2^3r^n},
\label{k0}
\end{equation*}
which gives the solution to the Third order Lovelock gravity as expected and for $\beta \rightarrow \infty ,$ $%
j(r)$ will be
\begin{equation}
j(r)=1-\frac{3\alpha_3}{2\alpha^2_2}
+\frac{3\alpha_{3}^{2}}{2\alpha_2^3}\left(\frac{1}{l^2}-\frac{(n-1)m}{r^{n}}-\frac{%
(n-2)e^{2}}{(n-4)r^{4}}\right). \label{jYM}
\end{equation}

\bigskip The solution given by Eqs. (\ref{f1}) and (\ref
{jYM}) represents the AdS solution to the Third order
Lovelock-Yang-Mills equation.

To see the asymptotic behavior of the solution to the Third order
Lovelock gravity, we write it for a special case that $\alpha
_{3}=\alpha _{2}^{2}.$ The solution then will be

\begin{eqnarray}
f(r)&=&-1+\frac{r ^{2}}{\alpha_{2}}+\frac{2^{1/3}r
^{2}}{\alpha_{2}}j(r)^{1/3}\label{ksp.} \\
j(r) &=&-\frac{1}{2}+\frac{3}{2}\alpha
_{2}\left\{\frac{1}{l^2}-\frac{(n-1)m}{r^{n}}+\frac{4\beta^2}{n(n-1)}\left[1-\digamma
(r)\right]\right\}. \notag
\end{eqnarray}

This solution has a singularity at $r=0$ as the kretschmann scalar
diverges at $r=0$.

The Hawking temperature for this solution is given by:

\begin{mathletters}
\begin{equation}
T=\frac{(n-2)r_+^3}{{(r_+^2-\alpha_2)}^{2}}\left(\Upsilon(r_+)
-\frac{(n-6)\alpha_2^2}{3(n-2)r_+^4}+\frac{(n-4)\alpha_2}{(n-2)r_+^2}
\right)\\
\end{equation}

where $r_{+}$ is the radius of event horizon. Also in this case,
we see that the black hole solutions may present an extreme black
hole with horizon radius $r_{\mathrm{ext}}$, where
$r_{\mathrm{ext}}$ is one of the real roots of $T=0$.

Now by solving the equations in second and third order Lovelock
theories, we deduce that the tt-component of the field equation in
Lovelock gravity is
\end{mathletters}
\begin{equation}
\sum_{i=0}^{p}\alpha _{i}^{^{\prime }}\frac{(n-2)!}{(n-2i)!%
}\frac{1}{nr^{n-3}}[r^{n}(\frac{-1-f(r)}{r^{2}})^{i}]^{^{\prime
}}+\frac{4\beta^2 r^2}{n(n-1)} \left(1-\Gamma\right)=0
\label{FieldLoven}
\end{equation}

 We can solve this equation to obtain
the general solution in Lovelock gravity as
\begin{equation}
\sum_{i=0}^{p}\alpha' _{i}\frac{(n-2)!}{(n-2i)!}(\frac{-1-f(r)}{%
r^{2}})^{i}=\frac{(n-1)m}{r^n}-\frac{4\beta^2}{n(n-1)}\left[1-\digamma
(r)\right].
\end{equation}

\section{Concluding Remarks}

We obtained the solutions to Einstein and Lovelock theories
considering the coupling of two non-linear fields, Yang Mills and
Born-Infeld fields and investigated the properties of the
solutions . The properties of these kind of solutions are like the
Yang-Mills solutions. But the difference seems to appear in the
role of the mass in the solutions, as for small r's in Yang-Mills
gravity the dominant term is the term containing $m$, but in the
the Yang-Mills-Born-Infeld, the dominant term indicates both $m$,
and $\digamma (r),$ which includes $\beta ,$ the Born-Infeld
parameter and $e$. As the function $\digamma (r),$ takes different
forms in different dimensions, we saw that this term modifies the
properties of the solutions. For example in Einstein-Yang-Mills
theory the solutions with nonnegative mass cannot present an
extreme black hole but we found conditions for some of the
solutions in Einstein-Yang-Mills-Born Infeld theory that, an
extreme black hole or naked singularity can exist for non-negative
mass.  The singularities in Einstein-Yang-Mills theory are always
spacelike and therefore unavoidable, but the singularities in
Einstein-Yang-Mills-Born-Infeld theory can be spacelike or
timelike depending on the choosing of the parameters. We also
obtained the solutions for the second and third order Lovelock
theories and from that we introduced the solution for the n-order
Lovelock-Yang-mills-Born-Infeld theory.\\

\textbf{Acknowledgements}

N. Bostani wishes to thank H. Hendi for many helps in this work.
This work has been supported in part by Research Institute for
Astrophysics and Astronomy of Maragha.

\end{document}